\documentclass[11pt]{emulateapj}
\usepackage{graphicx}
\usepackage{amssymb,amsmath}
\usepackage{natbib}
\begin{document}
\topmargin 0.5in 

\title{The Viewing Angles of Broad Absorption Line Versus Unabsorbed Quasars}
\author{M. A. DiPompeo\altaffilmark{1}, M. S. Brotherton\altaffilmark{1}, C. De Breuck\altaffilmark{2}}
\altaffiltext{1}{University of Wyoming, Dept. of Physics and Astronomy 3905, 1000 E. University, Laramie, WY 82071, USA}
\altaffiltext{2}{European Southern Observatory, Karl Schwarzschild Strasse 2, 85748 Garching bei M\"{u}nchen, Germany}

\begin{abstract}
It was recently shown that there is a significant difference in the radio spectral index distributions of broad absorption line (BAL) quasars and unabsorbed quasars, with an overabundance of BAL quasars with steeper radio spectra.  This result suggests that source orientation does play into the presence or absence of BAL features.  In this paper we provide more quantitative analysis of this result based on Monte-Carlo simulations.  While the relationship between viewing angle and spectral index does indeed contain a lot of scatter, the spectral index distributions are different enough to overcome that intrinsic variation.  Utilizing two different models of the relationship between spectral index and viewing angle, the simulations indicate that the difference in spectral index distributions can be explained by allowing BAL quasar viewing angles to extend about 10\arcdeg\ farther from the radio jet axis than non-BAL sources, though both can be seen at small angles.  These results show that orientation cannot be the only factor determining whether BAL features are present, but it does play a role.
\end{abstract}


\section{INTRODUCTION}
A long-time popular explanation for the presence of broad absorption lines (BALs) seen in approximately 20\% of quasar spectra (Knigge et al. 2008) has been a simple orientation model, in which BAL quasars are seen only from a more ``edge-on" perspective, or at larger viewing angles (Elvis 2000).  Understanding the geometry of the outflows producing these lines is an important part of modeling the role they play in the evolution of the quasar itself, as well as the effects they may have on the surrounding environment and host galaxy via feedback effects.  For example, it has been shown that it is possible for AGN feedback to affect star formation rates in the host galaxy, both theoretically (Hopkins \& Elvis 2010) and now it seems observationally (Cano-D\'{i}az et al. 2012).

The similarity of the emission lines in BAL and non-BAL quasars (Weymann et al. 1991), as well as their optical polarization properties (Ogle et al. 1999) have been used to support simple orientation models.  However, this scheme fails to explain various other observations, particularly at radio frequencies.  Short timescale radio variability has been identified in around 20 BAL quasars, which is argued to indicate a viewing angle near the radio jet axis to explain the derived brightness temperatures (Ghosh \& Punsly 2007, Zhou et al. 2006).  In general, BAL quasars are more compact than non-BALs in radio maps (Becker et al. 2000), at least at low to intermediate resolution, and in small samples they do not show a significant difference in radio spectral index distribution compared to non-BAL sources (Becker et al. 2000, Montenegro-Montes et al. 2008, Fine et al. 2011).  Radio spectral index ($\alpha$; $S_{\nu} \propto \nu^{\alpha}$, where $S_{\nu}$ is the radio flux and $\nu$ is the frequency) is generally considered an orientation indicator, with steeper spectrum ($\alpha < -0.5$) sources seen more edge-on because they are dominated by optically thin lobe emission, which has a steep spectrum due to little synchrotron self-absorption.  More pole-on sources are dominated by core emission because of relativistic beaming effects, and have a flatter spectrum because they are optically thick and thus significantly self-absorbed.  Due in large part to these observations, other explanations based on pure evolution (e.g. Gregg et al. 2006), and not orientation, have also come into favor.

DiPompeo et al. (2011) expanded greatly the number of BAL quasars with multi-frequency radio data and presented a sample of 74 BAL quasars, along with a sample of 74 individually matched unabsorbed quasars, with flux measurements at 4.9 and 8.4 GHz (observed frame) from the Very Large Array (VLA)/Expanded Very Large Array (EVLA).  These data provided quasi-simultaneous flux measurements to remove the effects of radio variability in the spectral index measurements.  The distributions of $\alpha_{8.4}^{4.9}$ were significantly different, with BAL quasars showing an overabundance of steep spectrum sources, but both samples show a wide range of spectral indices.  Analysis of other measurements of $\alpha$ (including $\alpha_{fit}$, a simple linear fit to available literature fluxes at various frequencies, in addition to the new measurements), a variety of statistical tests, and a restriction to only unresolved sources in both samples all also show that the difference is present, although the significance does vary.  Some of this variation could be due to the fact that the other measures of $\alpha$ included non-simultaneous flux measurements, or due to variation in the number of sources included in the tests.  Because the two samples are one-to-one matched in redshift, among other properties, use of rest-frame spectral indices will not effect the results.  

These results indicate that while BAL quasars likely span a range of orientations, viewing angle does plays a role in their presence.  The next step, presented here, is to quantify this difference and provide the most likely viewing angles to these sources, at least in a general sense.  Our aim is to test if the difference seen in spectral index distributions can be explained by differences in viewing angle, and we recognize that it may be possible to develop more sophisticated models and simulations in the future.

\section{THE $\alpha$-$\theta$ RELATIONSHIP}
We have based our modeling off of two relationships between $\alpha$ and viewing angle ($\theta$, defined as 0\arcdeg\ along the radio jet axis); one purely observational, and one from semi-empirical simulations.  We began developing the observational relationship with the sample of Wills \& Brotherton (1995), which included the 29 quasars of Ghisellini et al. (1993) with viewing angles calculated from superluminal motion seen in VLBI maps, as well as 4 additional sources from Vermeulen \& Cohen (1994) with superluminal motion measurements and X-ray data at 1 keV available.  We then searched NED\footnote{The NASA/IPAC Extragalactic Database (NED) is operated by the Jet Propulsion Laboratory, California Institute of Technology, under contract with the National Aeronautics and Space Administration} for multi-frequency radio fluxes in order to measure $\alpha$ for these sources.  We initially attempted to use fluxes at 4.85 and 8.4 GHz to better match the observations in DiPompeo et al. (2011), as well as a linear fit to all available radio fluxes, but given the small number of sources and the scatter in the $\alpha$-$\theta$ relationship we were unable to get a reasonable fit using these data.  One complication is that in general the radio flux measurements are not simultaneous, and variability almost certainly exaggerates the scatter since the majority of these sources are seen from small viewing angles.  In the end we settled on using the spectral index between 15 and 8.4 GHz, which was available for 27 of the 33 sources (26 from Ghisellini et al. (1993) and 1 from Vermeulen \& Cohen 1994), in order to build a useable model.  We note that a necessary assumption in this analysis is that on average the radio spectra are reasonably approximated by a power law over a large frequency range, which may be an oversimplification in some cases.  However, if you inspect the spectra presented in DiPompeo et al. (2011) for both BAL and non-BAL samples, when data is available at more frequencies the spectra are often well behaved across a large frequency range.

We adopt the values of the viewing angles from Ghisellini et al. (1993) and follow their method to calculate the viewing angle for the additional Vermeulen \& Cohen (1994) source.  Two sources that were large outliers (1830+285 and 1845+797, from Ghisellini et al. (1993), which both have highly inverted radio spectra) were excluded in order to allow us to make a fit to the data that could possibly reproduce our observed range of spectral indices.  The final list of sources, their viewing angles and their spectral indices are given in Table~\ref{modeldata}.  

\begin{figure*}
 \centering
  \figurenum{1}
   \includegraphics[width=5in]{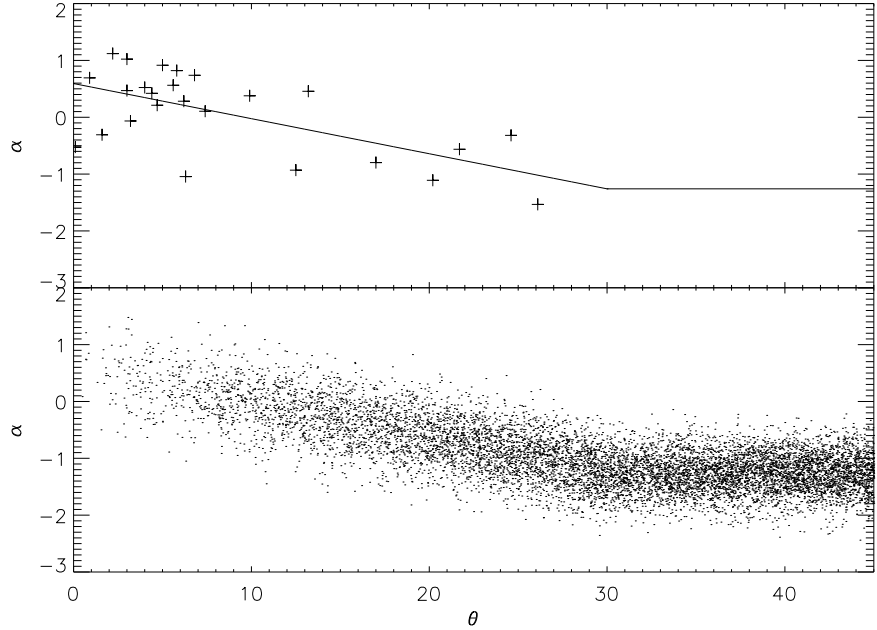}
  \caption{(Top) Observational data to constrain the $\alpha$-$\theta$ relationship, with the derived linear fit.  Here $\alpha$ is measured between 15 and 8.4 GHz due to constraints on available data, and we assume that this can be extended to lower frequencies.  Due to a lack of data and the fact that Doppler boosting does not likely effect the relationship above 30\arcdeg, we use a flat line there with some intrinsic scatter.  (Bottom)  Our monte-carlo simulation of 10,000 randomly distributed radio jets and their corresponding spectral index, determined by the observational data at the top.\label{obsfit}}
\end{figure*}

While the relationship is likely more complex in reality, we made a simple linear fit to the data.  Above a viewing angle of 30\arcdeg\ we assume a flat relationship where all the scatter is due to intrinsic variation in radio sources for several reasons.  First, there is no observational data above 30\arcdeg\ to constrain the shape of the relationship.  Second, if we simply extend the linear relationship, we would see large numbers of sources with extremely steep spectra ($\alpha$ extending to $-3$ or $-4$), which is simply not observed in most radio sources.  Finally, although it is dependent on the Lorentz factor of the the emitting material, the increase in observed flux due to Doppler boosting is expected to be small at larger viewing angles.  Because variability almost certainly exaggerates the scatter in the relationship, we use the standard deviation of the spectral index distribution of the quasars in the 3CR catalog (as presented in Smith \& Spinrad 1980) to model the scatter.  Because this catalog was built using low frequency data, it is dominated by lobe emission and therefore consists of mostly steep spectrum, edge-on sources.  Shown in the top of Figure~\ref{obsfit} is the data used from Ghisellini et al (1993) and Vermeulen \& Cohen (1994) and the fit used to model the $\alpha$-$\theta$ relationship; the bottom of Figure~\ref{obsfit} shows the result of our Monte-Carlo simulation of 10,000 randomly distributed jet viewing angles and their spectral indices based on the above model.

The semi-empirical model used is from the simulations of Wilman et. al (2008), which is the extragalactic portion of the SKA Simulated Skies (S$^3$) project.  We pulled from their simulations the fluxes at all available frequencies (151 MHz, 610 MHz, 1.4 GHz, 4.86 GHz, and 18 GHz, all observed frame) for all FR I and FR II type sources between redshifts of 1.5 and 3.5 and with values of $S_{1.4} \ge 10$ mJy.  Ideally we would use the spectral index between 4.86 and 18 GHz from these simulations; however, there are issues with their high frequency results.  The source populations are drawn from a 151 MHz luminosity function and extrapolated to high frequencies,  which almost certainly does not accurately predict the high frequency source population (see for example, Mahony et al. 2011).  Additionally, there may be problems with the functional form assumed for the SEDs of radio cores, causing them to fall off too steeply at high frequency (Wilman, private communication).  This causes the maximum spectral index found to drop well below 0 as you go to higher frequencies, which has no bearing on physical reality.  The simulations also apply a lower limit on the lobe spectral indices at $\alpha=-0.75$, which is problematic because we clearly see a high fraction of our sources with steeper spectra than this (the steep spectrum overabundance in BAL sources is most prominent between $-2 \le \alpha \le -1$).  However, if we again simply assume that in general radio spectra of these objects obey a simple power law, we can use the lower frequency spectral index distribution from these simulations and assume that a similar relationship holds between $\theta$ and $\alpha$ regardless of which part of the radio spectrum is considered.  Given the scatter in the relationship it is unlikely that this assumption will significantly effect our results.

\begin{figure*}
 \centering
  \figurenum{2}
   \includegraphics[width=5in]{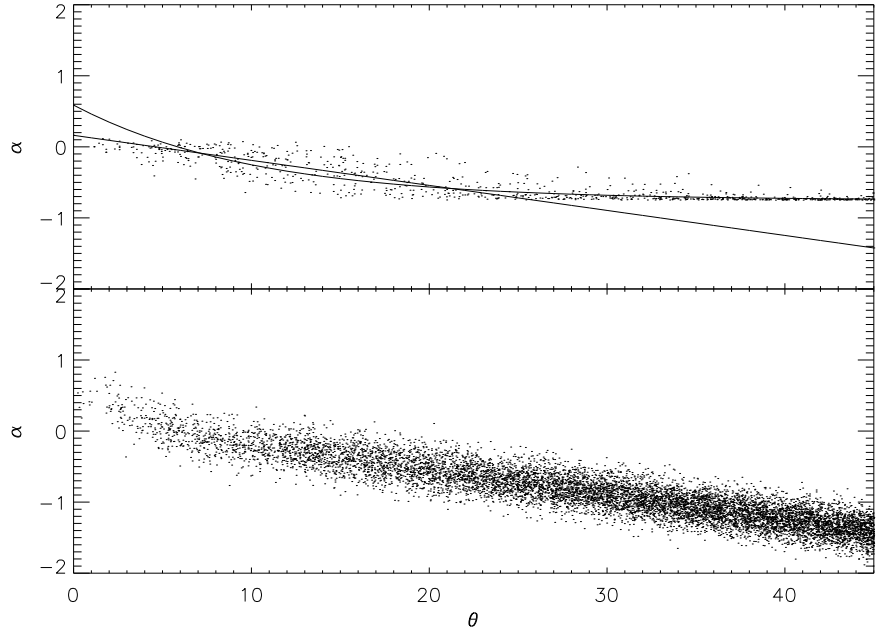}
  \caption{(Top) Simulation data from Wilman et al. (2008) showing the $\alpha$-$\theta$ relationship for all FRI and FRII sources between $1.5 \le z \le 3.5$ with $S_{1.4}$ greater than 10 mJy.  Here alpha is measured between 151 and 610 MHz, due to issues with the Wilman et al. (2008) simulations at higher frequency.  The linear fit is to the points between 10\arcdeg\ and 20\arcdeg\ before the data bottom out at $\alpha=-0.75$ (a lower limit used in the models) and flatten out at low $\theta$.  The curved line is a linear fit to $\log (\alpha+0.75)$ and $\theta$ between 5\arcdeg\ and 20\arcdeg.  Below 10\arcdeg\ the curved fit is used, above the linear fit is used.  (Bottom)  Our Monte-Carlo simulation of 10,000 randomly distributed radio jets and their corresponding spectral indicies, determined by the simulation data at the top.\label{simfit}}
\end{figure*}

The relationship between $\theta$ and $\alpha_{610}^{151}$ from these simulations is shown in the upper panel of Figure~\ref{simfit}.  In order to use this in our simulations, we made two fits to different ranges of $\theta$.  First, we did a simple linear fit from $10 \le \theta \le 20$, where the data can reasonably be approximated as linear.  This allows us to have sources with steeper spectra than $\alpha=-0.75$, which is needed to reproduce what is actually observed despite the limit applied in the simulations.  The distribution in $\alpha$ is then calculated by normalizing the data by the fit so all the variation is in $\alpha$, assuming the distribution is Gaussian and calculating the standard deviation.  The data is clearly non-linear at small $\theta$, and so between $5 \le \theta \le 20$ we also make a separate linear fit in $\log (\alpha + 0.75)$-$\theta$ space.  Our final model is then based on a combination of these fits; below viewing angles of 10\arcdeg\ the fit in logarithmic space is used, and above 10\arcdeg\ the linear fit is used.  Both lines are shown in the upper panel of Figure~\ref{simfit}.  The lower panel of Figure~\ref{simfit} shows the result of our Monte-Carlo simulation of 10,000 randomly distributed jet viewing angles and their spectral indices based on this semi-empirical model.

The overall general shape of the two models used is similar; however, the amount of scatter in the observational model is significantly more than that in the empirical model.  We also do not consider the possible effects of redshift in either model, though it is possible that the wide range of redshifts in the observed sample could have an effect on the spectral index distributions.  Again however, since the BAL and non-BAL samples are well matched in redshift, we feel that any effects of redshift will not change the general trends found. 

\section{SIMULATIONS}
The simulations were performed using IDL.  We first generate a random vector in 3-D space, utilizing IDL's uniformly distributed random number generator to create x, y, and z coordinates between -1 and 1, which are combined to create vectors within a ``unit cube''.  This will produce a distribution of vector directions biased toward the corners of the cube, and so we then reject any vector that does not fall within the unit sphere (has a magnitude greater than 1), resulting in vectors uniformly distributed in random directions in 3-D space.  This vector is taken to represent one side of a bipolar radio jet.

We next choose a viewing direction along the z-axis, and compute the viewing angle to the randomly generated vector.  If the vector is pointing in the negative z direction, we reflect it about the origin by multiplying all of its components by $-1$, to represent the opposite side of the bipolar jet, so that viewing angles range from 0 to 90\arcdeg.  Once the viewing angle is known, we can assign a value of $\alpha$ based on one of the models in the previous section.  This is done utilizing IDL's normally distributed random number generator, so that the assigned values of $\alpha$ are normally distributed about the model fit with the assumed standard deviation.  The bottom panels of Figures~\ref{obsfit} and~\ref{simfit} were generated by repeating this process $10^4$ times.

In order to compare the Monte-Carlo simulations to the real data and quantify the viewing angles to BAL and normal quasars, we used the simulation to virtually repeat our observational program while restricting what source viewing angles were allowed.  We systematically stepped through all possible ranges of allowed viewing angles between some $\theta_{min}$ and $\theta_{max}$, assigning values of $\alpha$ for each random viewing angle between the allowed $\theta_{min}$ and $\theta_{max}$ until there were 74 virtual ``observations'' (to compare to the 74 objects really observed in each sample).  We then compared the distribution of simulated spectral indices to the observed spectral index distributions for the BAL and non-BAL samples, using both K-S and R-S tests, to check whether the real data were well reproduced.  Our criterion was a p-value of less than 0.05 to indicate that the distributions are from a different parent population, and are not well matched.  We then repeat this experiment $10^5$ times, in order to get a statistical sense of how often the real observations are well reproduced for a given range of viewing angles.  Once this is complete, the allowed viewing angle range is changed and the process is repeated until all possible $\theta$ ranges are tested.  We apply an upper limit to $\theta_{max}$ of 45\arcdeg\, as above this value it is likely that most quasars are obscured by dust (Barthel 1989).  As the results will show in the next section, this upper limit has no effect on our findings.

\section{RESULTS}
\begin{figure*}
 \centering
  \figurenum{3}
   \includegraphics[width=6in]{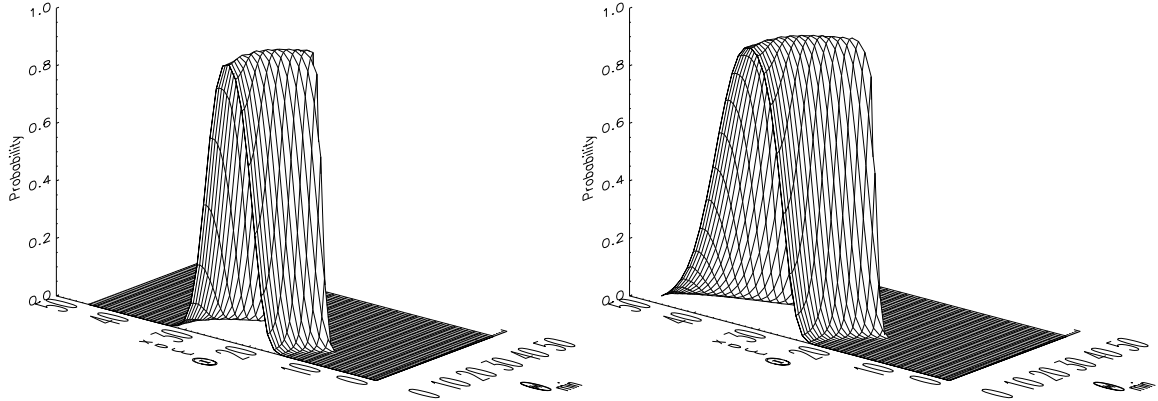}
  \caption{Comparison (via K-S tests) of simulations to observations of non-BALs (left) and BALs (right) for the observed $\alpha_{8.4}^{4.9}$, using the observationally constrained model.  The z-axis is the probability of the simulations matching the observations, as a function of $\theta_{min}$ (x-axis) and $\theta_{max}$ (y-axis).\label{resobscx}}
\end{figure*}

\begin{figure*}
 \centering
  \figurenum{4}
   \includegraphics[width=6in]{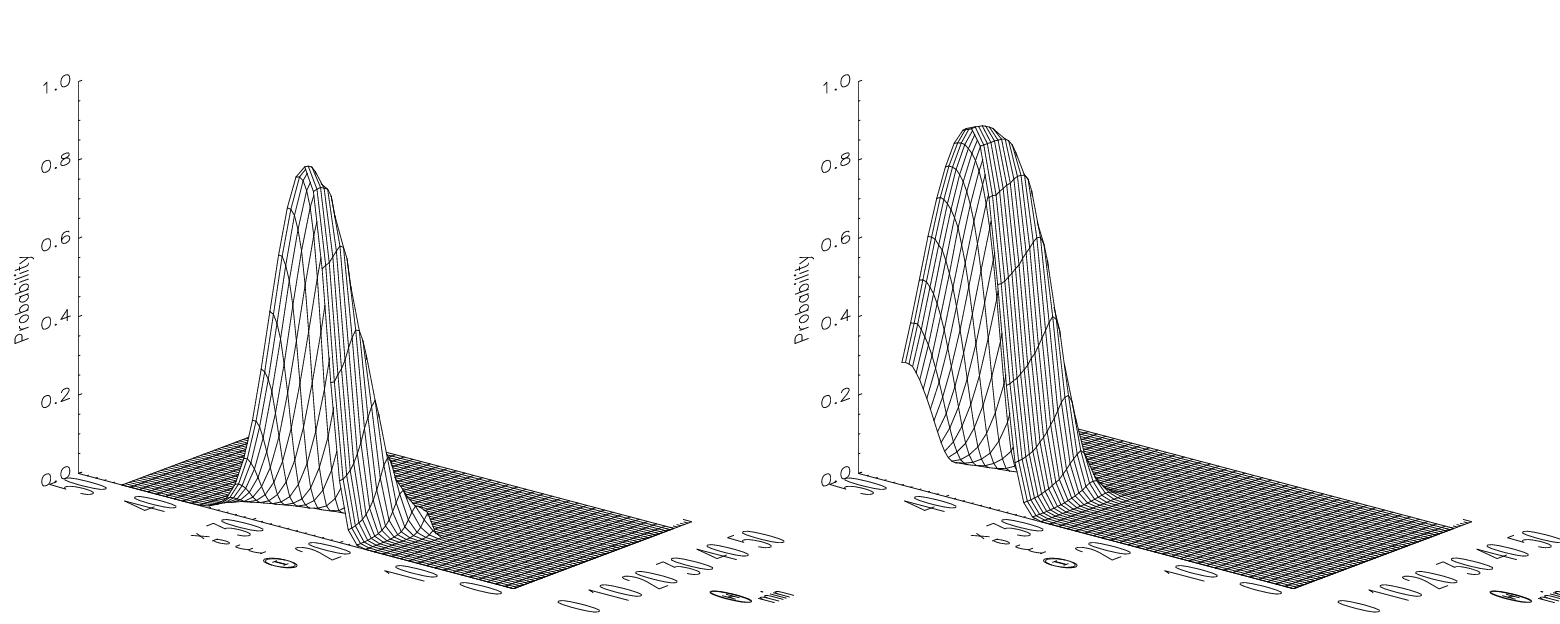}
  \caption{Comparison (via K-S tests) of simulations to observations of non-BALs (left) and BALs (right) for $\alpha_{8.4}^{4.9}$, using the semi-empirical model. The axes are the same as in Figure~\ref{resobscx}.\label{resmodcx}}
\end{figure*}

\begin{figure*}
 \centering
  \figurenum{5}
   \includegraphics[width=6in]{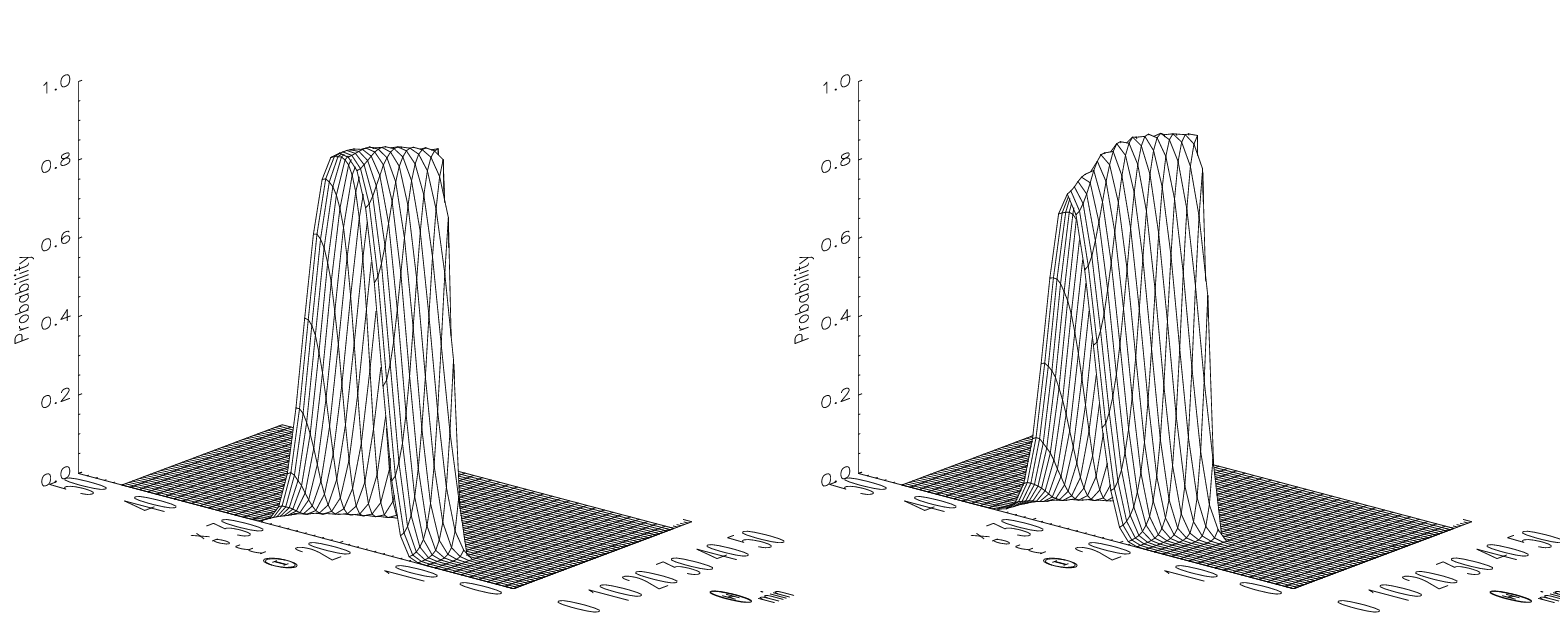}
  \caption{Comparison (via K-S tests) of simulations to observations of non-BALs (left) and BALs (right) for $\alpha_{fit}$, using the observational model. The axes are the same as in Figure~\ref{resobscx}.\label{resobsfit}}
\end{figure*}

\begin{figure*}
 \centering
  \figurenum{6}
   \includegraphics[width=6in]{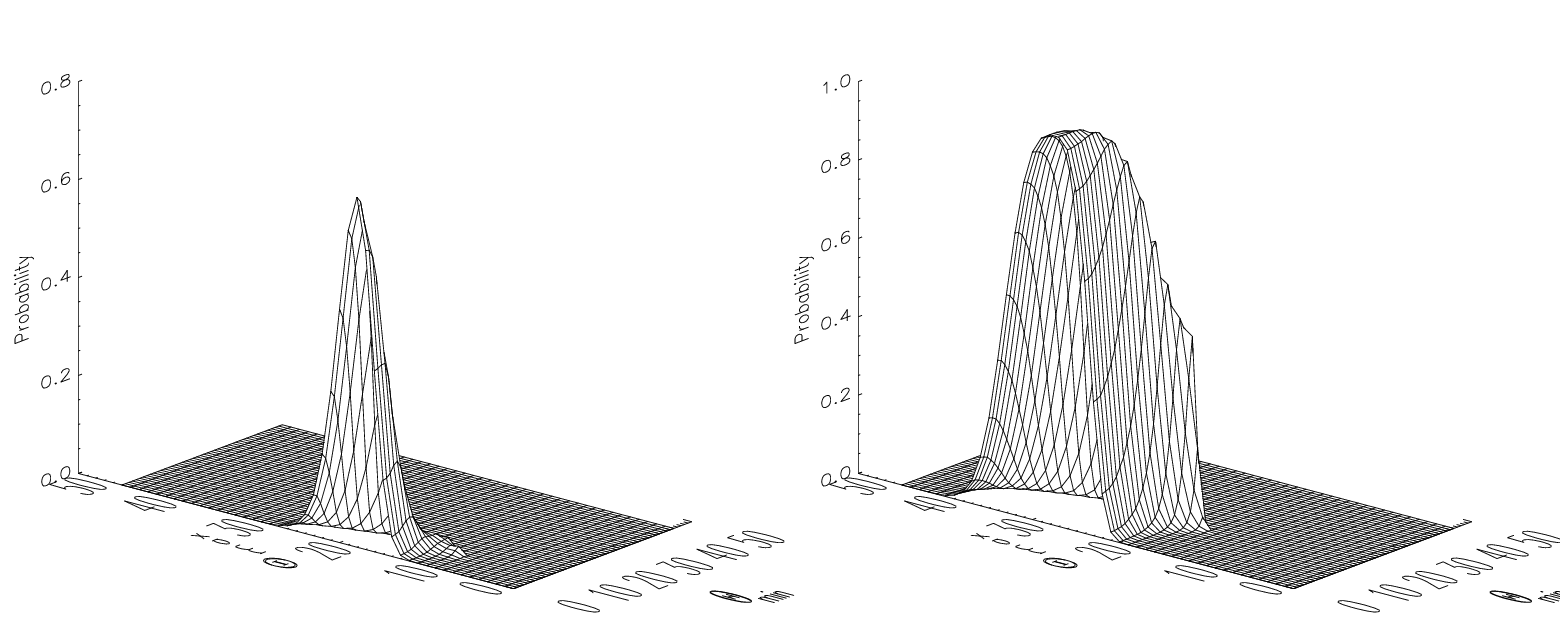}
  \caption{Comparison (via K-S tests) of simulations to observations of non-BALs (left) and BALs (right) for $\alpha_{fit}$, using the semi-empirical model.  The axes are the same as in Figure~\ref{resobscx}.\label{resmodfit}}
\end{figure*}

The probabilities of each set of simulations producing a match (based on the K-S tests) to the observed data as a function of $\theta_{min}$ and $\theta_{max}$ are shown in Figures~\ref{resobscx}-\ref{resmodfit}.  The z-axis represents the percentage (out of the $10^5$ runs) that the resulting $\alpha$ distribution matched the observed distribution, the x-axis is $\theta_{min}$ and the y-axis is $\theta_{max}$.  In each figure, the left panel shows the probabilities for the non-BAL sample, and the right panel shows the probabilities for the BAL sample.  Figures showing the probability distributions using an R-S test for comparison are not included to save space, since the results are not significantly different than using K-S tests.  The results are all summarized in Tables~\ref{resultstbl} (K-S test comparison) and~\ref{resultstbl2} (R-S test comparison); Column (1) is the sample that the simulations were compared to, Column (2) is the model used in the simulations (``Obs'' is the model based on the sample in Wills \& Brotherton (1995), ``Emp'' is the model based on the semi-empirical simulations of Wilman et al. (2008)), Column (3) is the maximum probability reached, Column (4) is the value of $\theta_{min}$ at $P_{max}$, and Column (5) is the value of $\theta_{max}$ at $P_{max}$.  The final two columns, (6) and (7) to the right of the vertical line, indicate the values of $\theta_{min}$ and $\theta_{max}$ with the largest separation that also have a probability of reproducing the observed results of greater than 90\%. 

In general, the simulations are able to accurately reproduce our observations, with more than half of them having probability distributions peaking at or above 99\%.  The only exception is trying to reproduce the $\alpha_{fit}$ distribution for non-BALs using the semi-empirical model and comparing using a K-S test, where the probability peaks at only 71\%.  However, the shape of the probability distribution and the location of the peak is still basically the same as other simulations.  The probability distributions are also generally quite flat in the $\theta_{min}$ direction, except for the cases where the semi-empirical model and a K-S test are used; in those situations the distributions are well-peaked.  This flatness is the reason we include the final two columns in Tables~\ref{resultstbl} and~\ref{resultstbl2}, as well as as the fact that the viewing angle ranges suggested by considering only the peaks can often be unrealistically narrow.  While there is some variation in the results depending on which model or statistical test is used, they are usually quite consistent.  In all cases, BAL quasars can have small viewing angles, but always extend farther from the radio jet axis when compared to non-BAL quasars.  If we take an average of all simulations, we find an average viewing angle range of 0-22\arcdeg\ for non-BALs and 1-32\arcdeg\ for BAL quasars.

\clearpage

\section{DISCUSSION}
With maybe the exception of the simulations using the observational model of the $\alpha$-$\theta$ relationship and comparing to $\alpha_{fit}$ with a K-S test, the general trend seen is that BAL quasars cover the same range of viewing angles as normal quasars, but extend about 10\arcdeg\ farther from the jet axis.  These results suggest that objects such as ``polar'' BALs (Ghosh \& Punsly 2007, Zhou et al. 2006) are indeed real and that for small to intermediate viewing angles it is possible to observe a quasar either with or without BALs, but at the largest viewing angles one will only see BAL features.  While constraining the $\alpha$-$\theta$ relationship can be difficult, it is interesting and encouraging that the results are quite similar using either of the two models or statistical comparisons here.  We suggest that more effort be placed on constraining this relationship (both observationally and theoretically) in the future, as sample sizes grow to significant enough numbers to use radio spectral index as a statistical orientation indicator when no other method is available.

One problem that is readily apparent in these results is that none of the samples extend to viewing angles as large as one might expect from, for example, the results of Barthel (1989).  This could indicate a problem with the models used.  However, the fact that the general result seems mostly independent of which model is used, we believe that while a change in the model may affect the absolute numbers it will not change the main conclusion.  Also, as mentioned in DiPompeo et al. (2011) there may be a slight biasing of our original sample toward more face-on sources (due to the requirement that all sources have 1.4 GHz FIRST fluxes greater than 10 mJy), but this bias should effect both BAL and non-BAL quasars equally.  So while it is possible that $\theta_{max}$ may be higher for both samples, the end result should remain the same.

Another consideration in this analysis is whether we can interpret the shape of the radio spectrum in a similar way in unresolved and extended radio sources.  To build models of the $\alpha$-$\theta$ relationship, we are required to utilize large scale, extended radio sources because these are the only sources in which $\theta$ can be measured directly, at least in sufficient numbers to build a useable model.  In contrast, the samples of DiPompeo et al. (2011) consist of high luminosity, generally compact radio sources (86\% and 78\% of BAL and non-BAL quasars, respectively, at 5\arcsec\ resolutions), although there is generally not data at high enough resolution or enough frequencies to classify them as belonging to the special classes of compact steep-spectrum (CSS) or especially gigahertz peaked-spectrum (GPS) sources.  It is possible that the shape of the spectrum in the samples is affected by evolution, as is the case for CSS sources, and not just geometry.  However, because CSS sources are selected to have steep spectra by definition, a comparison of spectral index distributions to test this theory would not be useful.  Instead, more radio data are needed, in particular at lower frequency, to look for turnovers in the radio spectra before comparisons with this class of object can be reasonably made.  

There are also indications that compact sources may have lower bulk velocities and thus lower Doppler factors (e.g., Polatidis \& Conway 2003).  If this is the case in these quasars, it is possible then that the core component is not enhanced at the same level as other sources and therefore the lobe component remains dominant to smaller viewing angles.  However, the fact that DiPompeo et al. (2011) still see a significant difference in spectral index distributions even when restricting the BAL and non-BAL samples to only unresolved sources suggests that there are still significant orientation effects.  Additionally, VLBI studies of BAL quasars (such as Doi et al. 2009, Kunert-Bajraszewska et al. 2010) do in fact show that compact BAL quasars have highly collimated radio jets, which means that viewing angle should effect the steepness of the radio spectrum in these sources in a similar way as more extended sources, as supported for example by the results of Jiang \& Wang (2003).  

Regardless of the complications mentioned above, these simulations should lay to rest any claims that BAL quasars are only seen edge-on, and move the argument away from simple dichotomies toward a more complex explanation that includes both orientation and other factors.  Of course with the data used here, we can only make these claims for radio-loud BAL quasars, though the similarities in other geometrically dependent properties between radio-loud and radio-quiet BAL quasars (for example, optical polarization; DiPompeo et al. 2010) suggest they may extend to radio-quiet BALs as well.  It is clear that to explain the spectral index distributions seen, there needs to be a large overlap in viewing angles to BAL and non-BAL quasars.  One simple explanation could be that BAL winds are launched at a variety of angles in different sources, or that they have a wide range of opening angles.  It is also possible that BALs can be explained by a combination of orientation and evolution.  The evolutionary schemes put forth for example by Gregg et al. (2002, 2006) suggest that BAL quasars begin enshrouded in a cocoon of gas and dust, which is blown out as radio jets develop and luminosity increases.  We may be able to draw on this description with these results, by making the clearing out of the gas and dust orientation dependent.  It is possible that the polar regions around the developing radio jets are cleared out first, which would explain the low numbers of seemingly ``polar'' BALs found as well as the spectral index distribution we observe.  The most equatorial regions then are never completely cleared of absorbing material (or are replenished by a disk wind), causing only BAL quasars to be seen beyond a particular viewing angle.  Whether a BAL is seen at small or intermediate angles is determined by the evolutionary status of the quasar, but in the latest stages of the quasar lifetime the presence of BALs is only orientation dependent.  We will attempt to develop this picture further in future papers, considering more factors than just the spectral index distributions modeled here.

\section{SUMMARY}
We have performed Monte-Carlo simulations of randomly oriented bi-polar radio jets and their corresponding radio spectral indices, in order to quantify the results of DiPompeo et al. (2011) that show a significant difference in the spectral index distributions of BAL versus non-BAL quasars.  The simulations utilize two different models of the relationship between spectral index and viewing angle; one based on observations and one based on semi-empirical simulations.  By limiting the allowed source viewing angles and comparing the resulting spectral index distributions with our observations, we can constrain the viewing angles to BAL and unabsorbed sources.  The results are mostly independent of which spectral index-viewing angle model is used, which spectral index measurement they are compared to, and which statistical test is used to compare the simulations to reality, at least in a general sense.  We find that there is a large overlap between viewing angles to the two samples, with both probably extending all the way down to 0\arcdeg\, or along the jet axis, which supports the claim that there are BAL quasars with polar outflows.  However, viewing angles to BAL sources generally extend farther from the jet axis, with about a 10\arcdeg\ span in which BALs will always be seen.

It is clear that a simple orientation-only model for BAL quasars cannot explain all of their observed properties, though orientation does play a role.  We need to move away from simple dichotomies in order to fully understand them, and it is likely that a combination of previous models will prevail in explaining this important subclass of quasar.

\acknowledgements
This research has made use of the NASA/IPAC Extragalactic Database (NED) which is operated by the Jet Propulsion Laboratory, California Institute of Technology, under contract with the National Aeronautics and Space Administration.  M. DiPompeo would like to thank the European Southern Observatory for funding visits to collaborate with C. De Breuck, as well as the Wyoming NASA Space Grant Consortium for providing funding to perform all of the radio observations.  M. DiPompeo would also like to thank Dr. Ruben Gamboa of the University of Wyoming Computer Science Department for assisting with the simulations, as well as Dr. Adam Myers for providing computing facilities to run the simulations.  Finally, we thank NRAO for their support in performing the radio observations that are the basis of these simulations; The National Radio Astronomy Observatory is a facility of the National Science Foundation operated under cooperative agreement by Associated Universities, Inc.

\clearpage

\begin{deluxetable}{ccc|ccc}
 \tabletypesize{\scriptsize}
 \tablewidth{0pt}
 \tablecaption{Sources Used in the Observational Model\label{modeldata}}
 \tablehead{
   \colhead{Source} & \colhead{$\theta$ ($\arcdeg$)}  & \colhead{$\alpha_{15}^{8.4}$} & \colhead{Source} & \colhead{$\theta$ ($\arcdeg$)}  & \colhead{$\alpha_{15}^{8.4}$} 
   }
   \startdata
    0016$+$731 & 5.0  &  0.91 & 1040$+$123 & 17.0 & -0.80 \\
    0106$+$013 & 3.0  &  0.47 & 1150$+$812 & 13.2 &  0.46 \\
    0153$+$744 & 26.1 & -1.53 & 1156$+$295 & 2.2  &  1.12 \\
    0212$+$735 & 6.2  &  0.28 & 1226$+$023 & 6.3  & -1.04 \\
    0333$+$321 & 3.0  &  1.02 & 1253$-$055 & 3.2  & -0.06 \\
    0430$+$052 & 9.9  &  0.37 & 1641$+$399 & 5.6  &  0.56 \\
    0552$+$398 & 20.2 & -1.11 & 1828$+$487\tablenotemark{a} & 6.8  &  0.74 \\
    0615$+$820 & 21.7 & -0.56 & 1928$+$738 & 7.4  &  0.11 \\
    0836$+$710 & 4.7  &  0.21 & 2134$+$004 & 0.1  & -0.53 \\
    0850$+$581 & 12.5 & -0.93 & 2223$-$052 & 1.6  & -0.31 \\
    0906$+$430 & 0.9  &  0.69 & 2230$+$114 & 4.0  &  0.52 \\
    0923$+$392 & 4.4  &  0.42 & 2251$+$158 & 5.8  &  0.82 \\
    1039$+$811 & 24.6 & -0.32 &            &      &	  \\
   \enddata
 \tablenotetext{a}{This source is from the sample of Vermeulen \& Cohen et al. (1994).  The viewing angle was calculated using the method of Ghisellini et al. (1993), using x-ray data from Wilkes et al. (1994) and additional radio data from Polatidis et al (1993).}
 \tablecomments{All sources and viewing angles are taken from Ghisellini et al. (1993) unless noted otherwise.  The spectral index $\alpha_{15}^{8.4}$ is measured from data gathered via NED.}
\end{deluxetable}

\begin{deluxetable}{cccccc|cc}
 \tablecaption{Simulation Results Using K-S Tests\label{resultstbl}}
 \tablehead{
   \colhead{Sample} & \colhead{Model} & \colhead{$\alpha$} & \colhead{$P_{max}$} & \colhead{$\theta_{min, P_{max}}$}  & \colhead{$\theta_{max, P_{max}}$}  & \colhead{$\theta_{min, P>0.9}$}  & \colhead{$\theta_{max, P>0.9}$}
  }
   \startdata
non-BAL   &   Obs    &   $\alpha_{8.4}^{4.9}$   &   0.99   &   12    &   19    &   0     &    24   \\[4pt]
BAL           &   Obs    &   $\alpha_{8.4}^{4.9}$   &   0.99   &   13    &   27    &   0     &    34   \\[4pt]
non-BAL   &  Emp    &  $\alpha_{8.4}^{4.9}$    &  0.93    &   0      &   24    &   0     &   25    \\[4pt]
BAL           &  Emp    &  $\alpha_{8.4}^{4.9}$    &  0.96    &   1      &   37    &   0     &   39    \\[4pt]
non-BAL   &   Obs    &   $\alpha_{fit}$               &   0.99   &   9       &   17    &   0     &    22   \\[4pt]
BAL           &   Obs    &   $\alpha_{fit}$               &   0.99   &   17     &   18    &   5     &    25    \\[4pt]
non-BAL   &  Emp    &  $\alpha_{fit}$                &  0.71   &   0        &  18     &  0\tablenotemark{a}      &   19\tablenotemark{a}   \\[4pt]
BAL           &  Emp    &  $\alpha_{fit}$                &  0.99   &   4        &   27    &  0      &    30
   \enddata
 \tablenotetext{a}{In this simulation the probability never reaches above 0.9, and so these values give the maximum range in $\theta$ where the probability is over 70\%}
 \tablecomments{See section 4 for a detailed explanation of the entries in this table.}
\end{deluxetable}

\begin{deluxetable}{cccccc|cc}
 \tablecaption{Simulation Results using R-S Tests\label{resultstbl2}}
 \tablehead{
   \colhead{Sample} & \colhead{Model} & \colhead{$\alpha$} & \colhead{$P_{max}$} & \colhead{$\theta_{min, P_{max}}$}  & \colhead{$\theta_{max, P_{max}}$}  & \colhead{$\theta_{min, P>0.9}$}  & \colhead{$\theta_{max, P>0.9}$}
  }
   \startdata
non-BAL   &   Obs    &   $\alpha_{8.4}^{4.9}$   &   0.99   &   15    &   17    &   0     &    24   \\[4pt]
BAL           &   Obs    &   $\alpha_{8.4}^{4.9}$   &   0.99   &   18    &   24    &   0     &    34   \\[4pt]
non-BAL   &  Emp    &  $\alpha_{8.4}^{4.9}$    &  0.99    &   14      &   17    &   0     &   25    \\[4pt]
BAL           &  Emp    &  $\alpha_{8.4}^{4.9}$    &  1.00    &   21      &   27    &   0     &   40    \\[4pt]
non-BAL   &   Obs    &   $\alpha_{fit}$               &   0.98   &   13       &   14    &   0     &    21   \\[4pt]
BAL           &   Obs    &   $\alpha_{fit}$               &   0.97   &   15     &   20      &   0     &    27    \\[4pt]
non-BAL   &  Emp    &  $\alpha_{fit}$                &  1.00   &   7        &  14     &  0      &   21   \\[4pt]
BAL           &  Emp    &  $\alpha_{fit}$                &  0.99   &   17        &   28    &  0      &    29  
   \enddata
 \tablecomments{See section 4 for a detailed explanation of the entries in this table.}
\end{deluxetable}

\clearpage


\clearpage





\begin{thebibliography}

\bibitem[Barthel (1989)]{Barthel1989}  Barthel, P.D.  1989, \apj, 336, 606

\bibitem[Becker (2000)]{Becker2000} Becker, R. H., White, R. L., Gregg, M.D., Brotherton, M.S., Laurent-Muehleisen, S.A. \& Arav, N.  2000, \apj, 538, 72

\bibitem[DiPompeo (2011)]{DiPompeo2011} DiPompeo, M.A., Brotherton, M.S., De Breuck, C., Laurent-Muehleisen, S.A.  2011, \apj, 473, 71

\bibitem[Doi (2009)]{Doi2009} Doi, A., Kawaguchi, N, Kono, Y. et al.  2009, PASJ, 61, 1389

\bibitem[Elvis (2000)]{Elvis2000} Elvis, M.  2000, \apj, 545, 63

\bibitem[Fine (2011)]{Fine2011} Fine, S., Jarvis, M.J., \& Mauch, T.  2011, \mnras, 412, 213

\bibitem[Ghosh (2007)]{Ghosh2007} Ghosh, K.K. \& Punsly, B.  2007, \apj, 661, L139

\bibitem[Ghisellini (1993)]{Ghisellini1993} Ghisellini, G., Padovani, P., Celotti, A., \& Maraschi, L.  1993, \apj, 407, 65

\bibitem[Gregg (2006)]{Gregg2006} Gregg, M.D., Becker, R.H. \& de Vries, W.  2006, \apj, 641, 210

\bibitem[Gregg (2002)]{Gregg2002} Gregg, M.D., Lacy, M., White, R.L., Glikman, E., Helfand, D., Becker, R.H., Brotherton, M.S.  2002, \apj, 564, 133

\bibitem[Hopkins (2010)]{Hopkins2010} Hopkins, P.F. \& Elvis, M.  2010, \mnras, 401, 7

\bibitem[Jiang (2003)]{Jiang2003} Jiang, D.R., \& Wang, T.G.  2003, A\&A, 397, L13

\bibitem[Knigge (2008)]{Knig2008} Knigge, C., Scaringi, S., Goad, M.R. \& Cottis, C.E.  2008, \mnras, 386, 1426

\bibitem[Kunert (2010)]{Kunert2010} Kunert-Bajraszewska, M., Jankiuk, A., Gawronski, M.P., \& Siemiginowska, A.  2010, \apj, 718, 1345

\bibitem[Montenegro (2008)]{MM2008} Montenegro-Montes, F.M., Mack, K.-H., Vigotti, M., Benn, C.R., Carballo, R., Gonzalez-Serrano, J.I., Holt, J. \& Jimenez-Lujan, F.  2008, \mnras, 388, 1853

\bibitem[Mahony (2011)]{Mahony}  Mahony, E.K., Sadler, E.M., Croom, S.M., Ekers, R.D., Bannister, K.W., Chhetri, R., Hancock, P.J., Johnston, H.M., Massardi, M., \& Murphy, T.  2011, \mnras, 417, 2651

\bibitem[Ogle (1999)]{Ogle1999} Ogle, P.M., Cohen, M.H., Miller, J.S., Tran, H.D., Goodrich, R.W., 

\bibitem[Polatidis (2003)]{Polatidis2003} Polatidis, A.G., \& Conway, J.E.  2003, PASA, 20, 69

\bibitem[Polatidis (1993)]{Polatidis1993} Polatidis, A.G., Wilkinson, P.N., Akujor, C.E.  1993, Sub-arcsecond Radio Astronomy, Proceedings of the Nuffield Radio Astronomy Laboratories' Conference, p.225

\bibitem[Smith (1980)]{Smith1980} Smith, H.E. \& Spinrad, H.  1980, ASP, 92, 553

\bibitem[Vermeulen (1994)]{Verm1994} Vermeulen, R.C., \& Cohen, M.H.  1994, \apj, 430, 467

\bibitem[Weymann (1991)]{Weymann1991} Weymann, R.J. \& Morris, S.L.  1991, \apj, 373, 23

\bibitem[Wilman (2008)]{Wilman2008} Wilman, R.J., Miller, L., Jarvis, M.J., Mauch, T., Levrier, F., Abdalla, F.B., Rawlings, S., Klockner, H.-R., Obreschkow, D., Olteanu, D., \& Young, S.  2008, \mnras, 388, 1335

\bibitem[Wilkes (1994)]{Wilkes1994} Wilkes, B.J., Tananbaum, H., Worral, D.M., Avni, Y., Oey, M.S., Flanagan, J.  1994, \apjs, 92, 53

\bibitem[Wills (1995)]{Wills1995} Wills, B.J. \& Brotherton, M.S.  1995, \apjl, 448, L81

\bibitem[Zhou (2006)]{Zhou2006} Zhou, H., Want, T., Wang, H., Wang, J., Yuan, W. \& Lu, Y.  2006, \apj, 639, 716



\end{thebibliography}
\end{document}